\documentclass[journal]{IEEEtran}
\usepackage{color}
\usepackage[pdftex]{graphicx}
\usepackage{graphicx}
\usepackage{caption}
\usepackage{subcaption}
\newcommand{\protovipram}{{\tt protoVIPRAM00}}
\begin{document}

\title{Performance Study of the First 2D Prototype of Vertically Integrated Pattern Recognition Associative Memory (VIPRAM)}

\author{
\IEEEauthorblockN{Gregory Deptuch, James Hoff, Sergo Jindariani, Tiehui Liu, Jamieson Olsen, Nhan Tran\\}
\IEEEauthorblockA{ {\it Fermi National Accelerator Laboratory, Batavia, IL 60510, USA} \\ }
\and 
\vspace{0.3cm}
\IEEEauthorblockN{Siddhartha Joshi, Dawei Li, Seda Ogrenci-Memik\\}
\IEEEauthorblockA{ {\it Department of EECS, Northwestern University, Evanston, IL 60201, USA} \\}
}

\maketitle

\begin{abstract}
Extremely fast pattern recognition capabilities are necessary to find and fit billions of tracks at the hardware trigger level produced every second  anticipated at high luminosity LHC (HL-LHC) running conditions. 
Associative Memory (AM) based approaches for fast pattern recognition have been proposed as a potential solution to the tracking trigger. 
However, at the HL-LHC, there is much less time available and speed performance must be improved over previous systems while maintaining a comparable number of patterns. 
The Vertically Integrated Pattern Recognition Associative Memory (VIPRAM) Project aims to achieve the target pattern density and performance goal using 3DIC technology. 
The first step taken in the VIPRAM work was the development of a 2D prototype (\protovipram) in which the associative memory building blocks were designed to be compatible with the 3D integration. 
In this paper, we present the results from extensive performance studies of the \protovipram~chip in both realistic HL-LHC and extreme conditions. 
Results indicate that the chip operates at the design frequency of 100 MHz with perfect correctness in realistic conditions and conclude that the building blocks are ready for 3D stacking.
We also present performance boundary characterization of the chip under extreme conditions.
\end{abstract}


%
\IEEEpeerreviewmaketitle


\section{Introduction}
\label{sec:intro}

Associative memory (AM)-based pattern recognition is a powerful approach to solving complex combinatorics for fast triggering on particle tracks~\cite{Dell'Orso:1988zz}.
It has been successfully used in previous high-energy physics experiments 
such as the CDF Silicon Vertex Trigger (SVT) at the Fermilab Tevatron~\cite{2004NIMPA.518..532A} 
and is currently being implemented at the ATLAS FastTracKer (FTK) at the CERN LHC~\cite{ftkTDR}. 
This massively parallel architecture is ideally suited to tackle the intrinsic combinatorics of track finding algorithms, 
avoiding the typical power-law dependence of execution time on occupancy and solving the pattern recognition in times roughly proportional to the number of hits. 
This is of crucial importance given the large occupancies typical of hadronic collisions and low latency requirements. 

There will be a much higher hit occupancy than ever seen before in proton-proton collisions at the anticipated high luminosity Large Hadron Collider (HL-LHC) at CERN.
This effect, typically called pileup, comes from multiple (140-200) collisions per proton bunch collision.
In order to keep the rate of events manageable for the experiments, a track trigger system will be required~\cite{cmsTP}.
The latency requirements of such a track trigger system require it to be operated at speeds much faster than previous systems.
At the same time, the HL-LHC detectors are being designed to have a much larger number of channels in their tracking volume than previous LHC and 
Tevatron detectors, and thus, there is an enormous challenge in implementing fast pattern recognition for a track trigger.

The rigorous technical requirements of a silicon-based hardware tracking trigger push the limits of 
Pattern Recognition Associative Memories (PRAM) in pattern density, speed, and power density.
It is estimated that the number of patterns needed for a tracking trigger system at CMS~\cite{cmsTP} for the HL-LHC is two orders of magnitude greater than the CDF system~\cite{2004NIMPA.518..532A} requiring a greatly increased pattern density.
For a similar system, the latency estimate is at least an order of magnitude less than the ATLAS FTK system~\cite{ftkTDR} requiring greatly increased speed performance. 

The associative memory approach involves using content addressable memories (CAMs) and a majority logic (ML) to find matching detector hits from different detector layers to form track candidates with extremely low latencies. 
Approaches to this goal in simple 2D VLSI, which were previously used, are limited.  
Reducing the feature size of the technology node while scaling up the number of patterns is an option~\cite{amchip05,amchip06,wetext},
but can present its own design challenges.  
Given these new challenges, a new concept to use emerging 3D technology has been proposed~\cite{Liu:2011zzw}.
Design in 3D vertical integration is, in a sense, the logical partitioning of functionality into a third dimension. 
The PRAM structure is intrinsically adaptable to the 3rd dimension from the full pattern level down to the individual CAM level. 

\begin{figure}
\centering
	\begin{subfigure}[bht]{0.35\textwidth}
	\includegraphics[width=\textwidth]{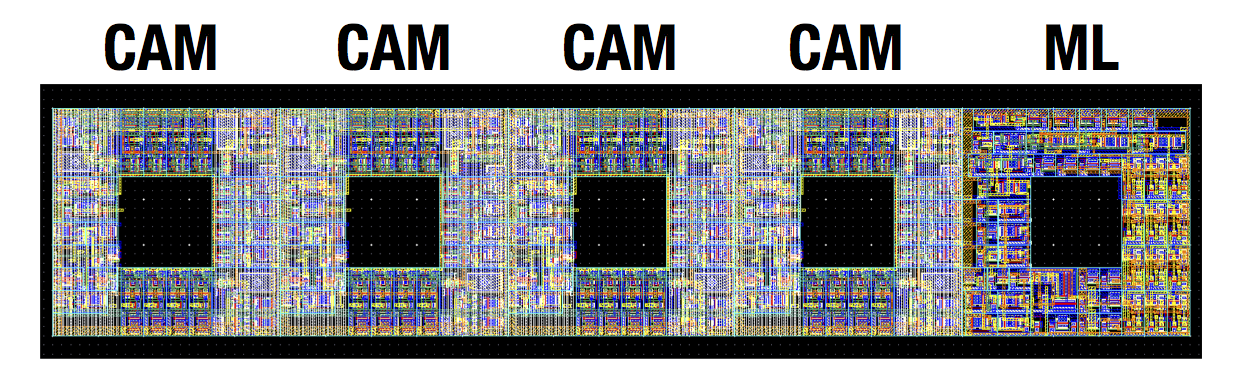}
	\caption{}
	\end{subfigure}
\vspace{0.25cm}\\
	\begin{subfigure}[bht]{0.2\textwidth}
	\includegraphics[width=\textwidth]{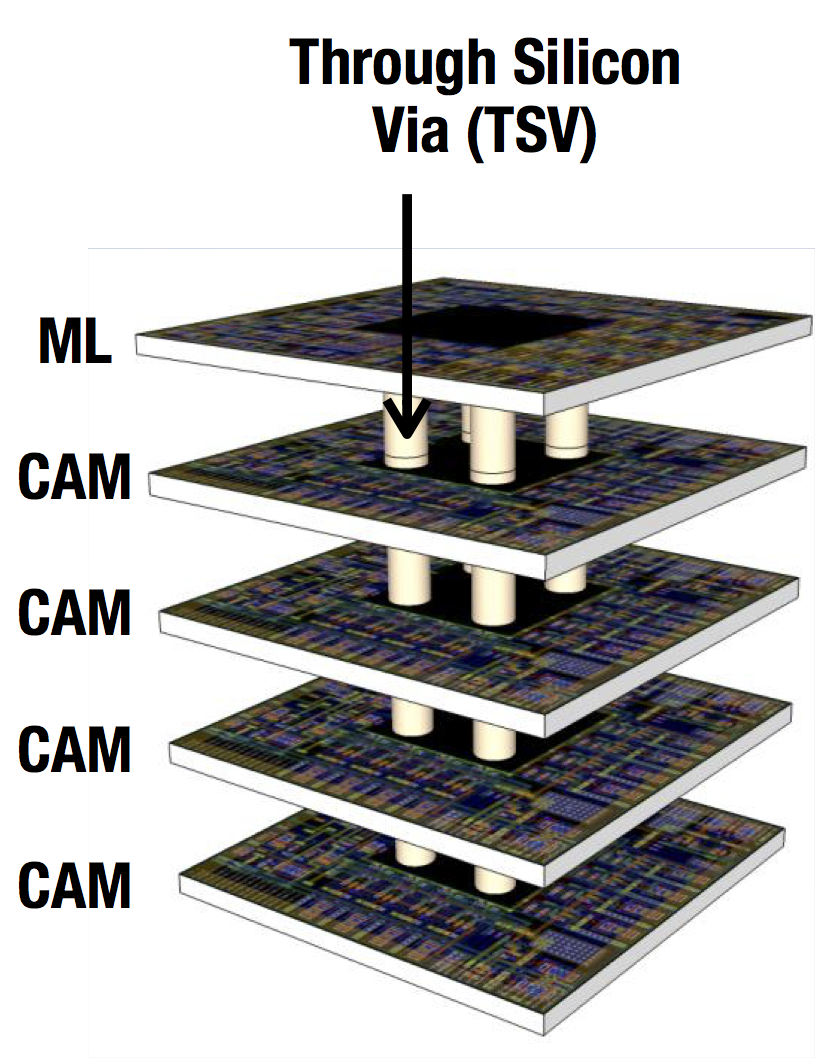}
	\caption{}
	\end{subfigure}
	\begin{subfigure}[bht]{0.2\textwidth}
	\includegraphics[width=\textwidth]{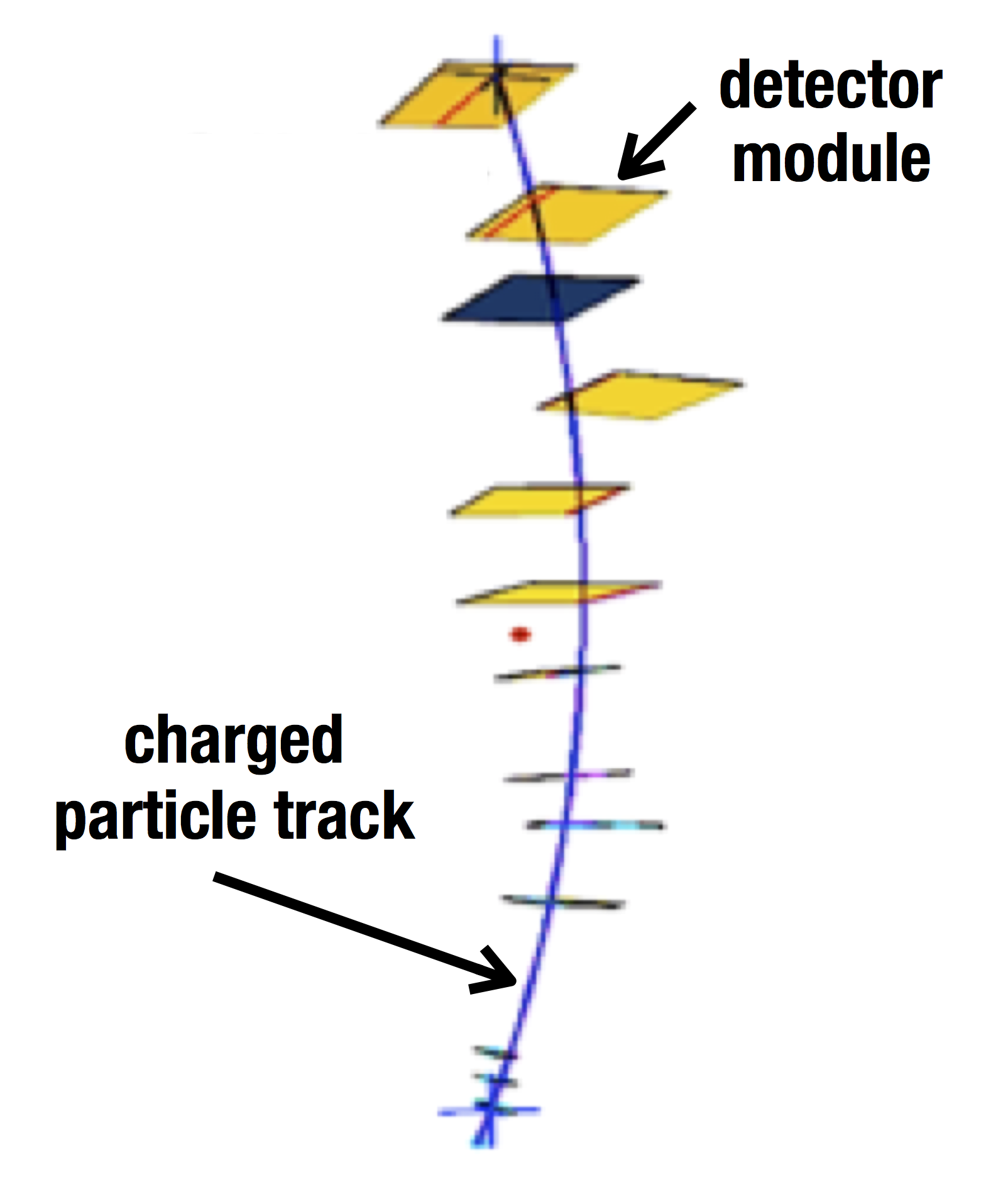}
	\caption{}
	\end{subfigure}
\caption{Pattern Recognition Associative Memory laid out in 2D (a) and stacked in 3D (b).  An illustration of the signal for a charged particle in detector layers is given in (c).}
\label{fig:stack}
\end{figure}

The VIPRAM (Vertically Integrated PRAM) approach is to divide the PRAM structure among 3D VLSI tiers to reduce the area consumed by a single pattern, 
to reduce the parasitic capacitance of long traces, to increase the effective number of routing layers, and to increase the readout speed significantly. 
The essence of VIPRAM is to divide this approach up into different tiers, maximizing pattern density while minimizing critical lengths and parasitics and therefore the power density.
In Fig.~\ref{fig:stack}(a), a PRAM element is laid out in 2D VLSI and in Fig.~\ref{fig:stack}(b), 3D VLSI on the bottom.
In Fig.~\ref{fig:stack}(c), a charged particle which represents a pattern is illustrated where each location in the detector can be associated with a CAM cell.
From the figure, one can see that the pattern density directly depends on the cross-sectional area of one of the CAM cells and is greatly increased by stacking each of the PRAM components.
The lines from the CAM to the majority logic (ML) cell, which contains logic to assert a match, are long in the conventional 2D implementation. 
They are now implemented vertically and are therefore shorter in 3D because each tier will be thinned down to about 10 $\mu m$ during the 3D stacking process. 
Due to the high number of repetitive structures in associative memories, 3D integration can have a significant impact on performance. 
The vertical integration also provides flexibility in layout optimization of the building blocks, and therefore chip performance.

We present details related to design and testing of a 2D prototype of the 3D VIPRAM concept, which we refer to as \protovipram.
Because vertical integration is an emerging technology, we have studied first the basic building blocks of the 3D concept laid out in 2D 
to verify their performance in simple 2D VLSI.  
Then, in the next version of the chip which includes 3D integration, we will be able to directly compare and quantify the expected performance gains with respect to the 2D layout.

Previous discussion on the chip details, design, simulation, and initial test results can be found in~\cite{Liu:2015oca}. 
The main focus of this paper is detailed testing results.
In Section~\ref{sec:design}, we describe the design and functionality of the \protovipram~from the single PRAM design to the full prototype chip layout.
In Section~\ref{sec:testbench}, the testing setup for the \protovipram~chips is detailed.  
This is comprised of a simple setup to test the generic functionality of the chip and also a more sophisticated setup for testing realistic scenarios and boundary conditions.
In Section~\ref{sec:results}, the results of the \protovipram~testing are presented.
First, we describe the tests of the basic functionality of the chip as a generic tool for pattern recognition.
Second, scenarios based on realistic HL-LHC simulations are presented.
Third, we detail tests which are designed to test extreme boundary conditions of the chip.  
Finally, for the various tests performed, we examine the corresponding power consumption which will be important in contrasting and benchmarking future versions of the chip.
In Section~\ref{sec:concl}, we summarize the paper and provide an outlook on the project status.

\section{Associative memory building blocks}
\label{sec:design}

Since 3D Vertical Integration is an emerging technology and the requirements of the hardware track trigger have themselves been evolving, 
the first logical step is to test the two basic building blocks, the CAM cell and the majority logic cell, through a simple 2D prototype run. 
This will provide verification of their functionality in preparation for the 3D stacking and low latency readout developments in the near future.
The associative memory building blocks were laid out as if this was a 3D design.
Space was reserved for as yet non-existent through silicon vias (TSV) and routing was performed to avoid these areas.
The readout circuitry of the PRAM array is deliberately simplified to allow for direct performance studies of the CAM and control cells.
The 2D prototype run also serves as a benchmark to understand the performance improvements that can be gained by 3D stacking.

The \protovipram~was designed and fabricated in a 130~nm Low Power CMOS process that has been used previously in HEP 3D designs. 
Fabrication is performed by Global Foundaries.  
The size of the chip is 5.46~mm $\times$ 5.46~mm. The layout was implemented such that, in future 3D designs, the basic building blocks can be directly reused and placed on different 3D tiers.
The prototype chip has 4096 patterns distributed in 128 rows and 32 columns.

\subsection{Single PRAM design}

\begin{figure}[!t]
\centering
	\begin{subfigure}[bht]{0.35\textwidth}
	\includegraphics[width=\textwidth]{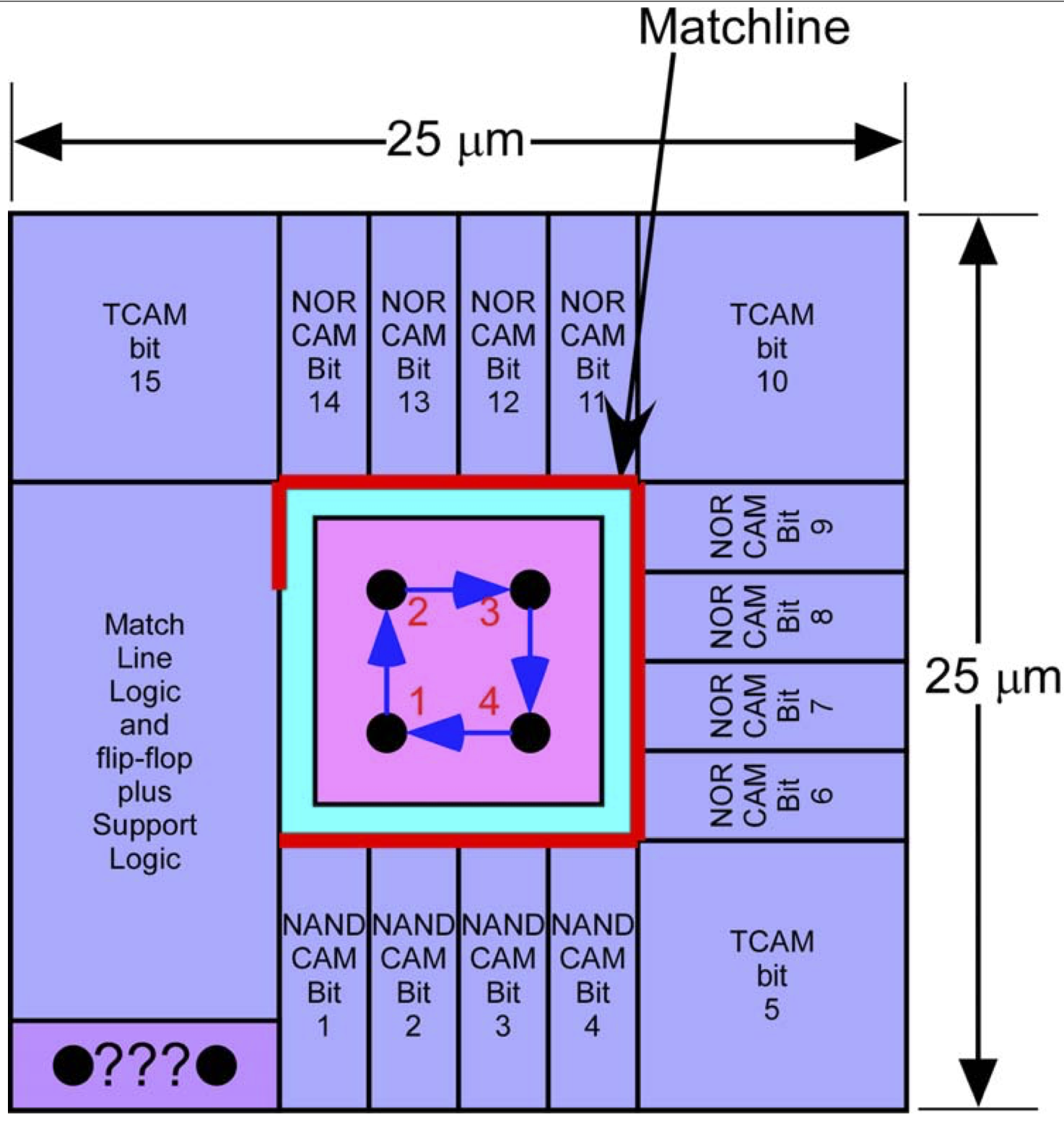}
	\caption{}
	\end{subfigure}
\vspace{0.25cm}
	\begin{subfigure}[bht]{0.35\textwidth}
	\includegraphics[width=\textwidth]{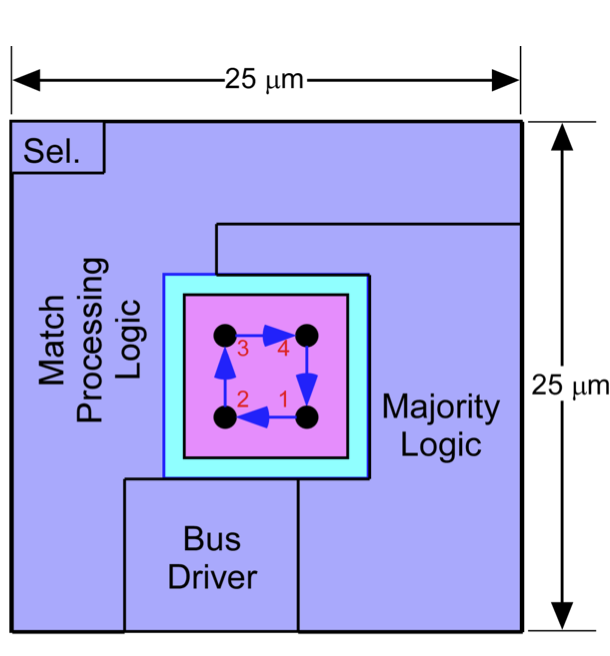}
	\caption{}
	\end{subfigure}
\caption{CAM Cell schematic (a) and Majority Logic Cell (b).}
\label{fig:PRAM}
\end{figure}

In keeping with the design philosophy of testability, each PRAM pattern consists of four identical CAM cells 
and a control cell resulting in the ability to recognize 4-layer pattern matches.
The choice of four CAM cells is made for simplicity though realistic systems will require more where current designs consider eight.  
Each CAM cell is a 15-bit address where the 15-bits are comprised of 4 NAND cells, 8 NOR cells, and 3 Ternary bits with a 4-bit selective pre-charge~\cite{Liu:2011zzw}.
Figure~\ref{fig:PRAM}(a) shows the floor plan of the CAM cell including the space in the middle reserved for TSVs.
The {\it Matchline}, indicated in Fig.~\ref{fig:PRAM}(a), is the single signal that connects the different bits in the CAM cell and its parasitic impedance impacts the chip performance.
The selective pre-charge is made with four NAND cells which, when matched, allow the Matchline to be charged.
The choice of the number of pre-charge bits was made to optimize performance and is a balance between less power consumption (more pre-charge bits) and increased clock frequency (fewer pre-charge bits).

The Majority Logic cell, shown in Fig.~\ref{fig:PRAM}(b), is designed to have the same footprint as the CAM cell.
The Majority Logic uses Pass Transistor Logic to produce a 3-bit code indicative of the possible match conditions: All Layer Match, One Missing Layer Match, Two Missing Layers Match, and First Layer Match. 
The Match Processing Logic compares the output of the Majority Logic with the user-supplied threshold and, if met, asserts a matched pattern.

Each of the majority logic and CAM cells are 25$\mu$m $\times$ 25$\mu$m in size and there is an additional 10$\mu$m $\times$ 125$\mu$m of space left for pattern routing
such that a PRAM is 35$\mu$m $\times$ 125$\mu$m. 

\subsection{Full \protovipram~design}
\label{sec:fullchip}

\begin{figure}[!t]
\centering
	\begin{subfigure}[bht]{0.40\textwidth}
	\includegraphics[width=\textwidth]{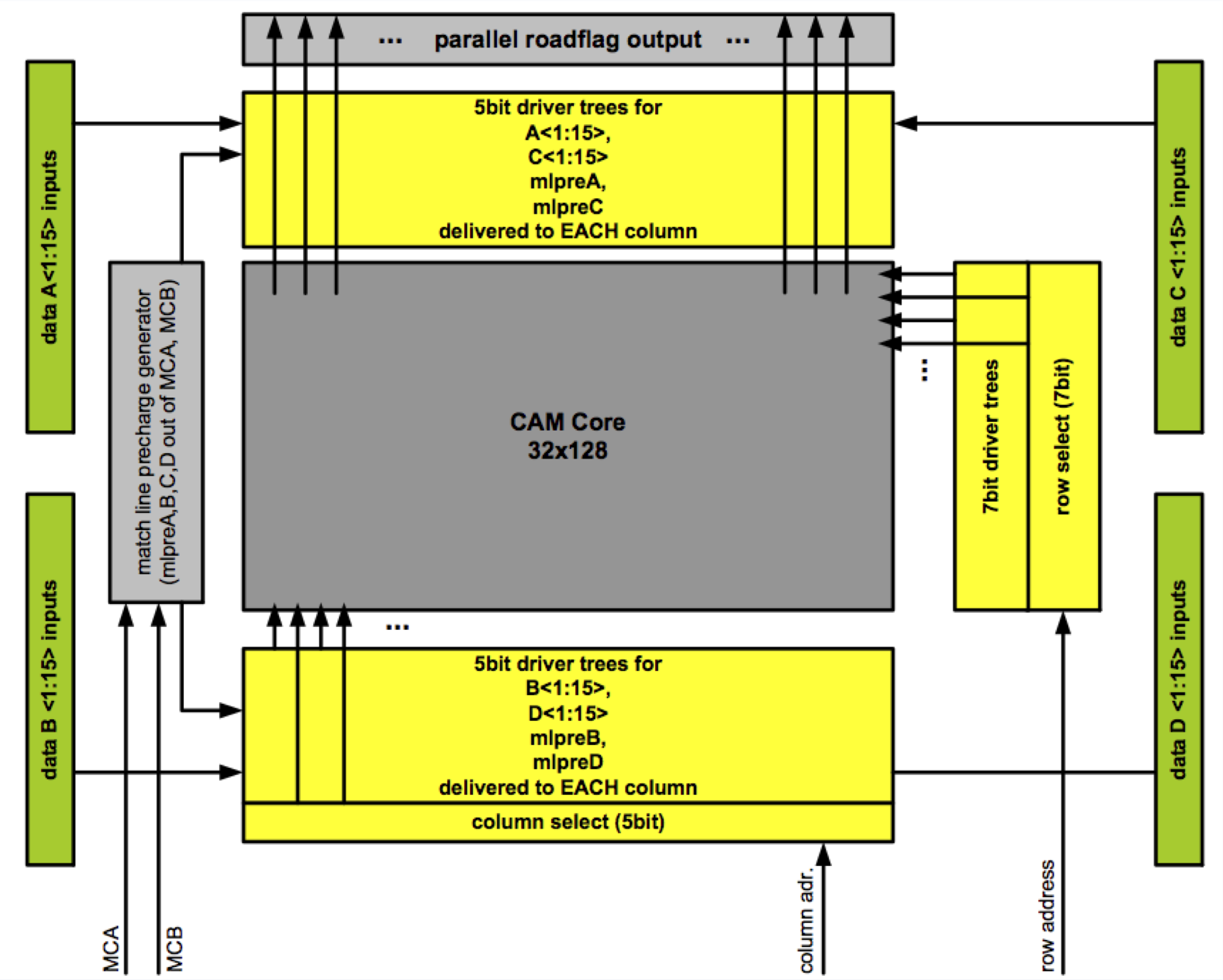}
	\caption{}
	\end{subfigure}
\vspace{0.25cm}
	\begin{subfigure}[bht]{0.35\textwidth}
	\includegraphics[width=\textwidth]{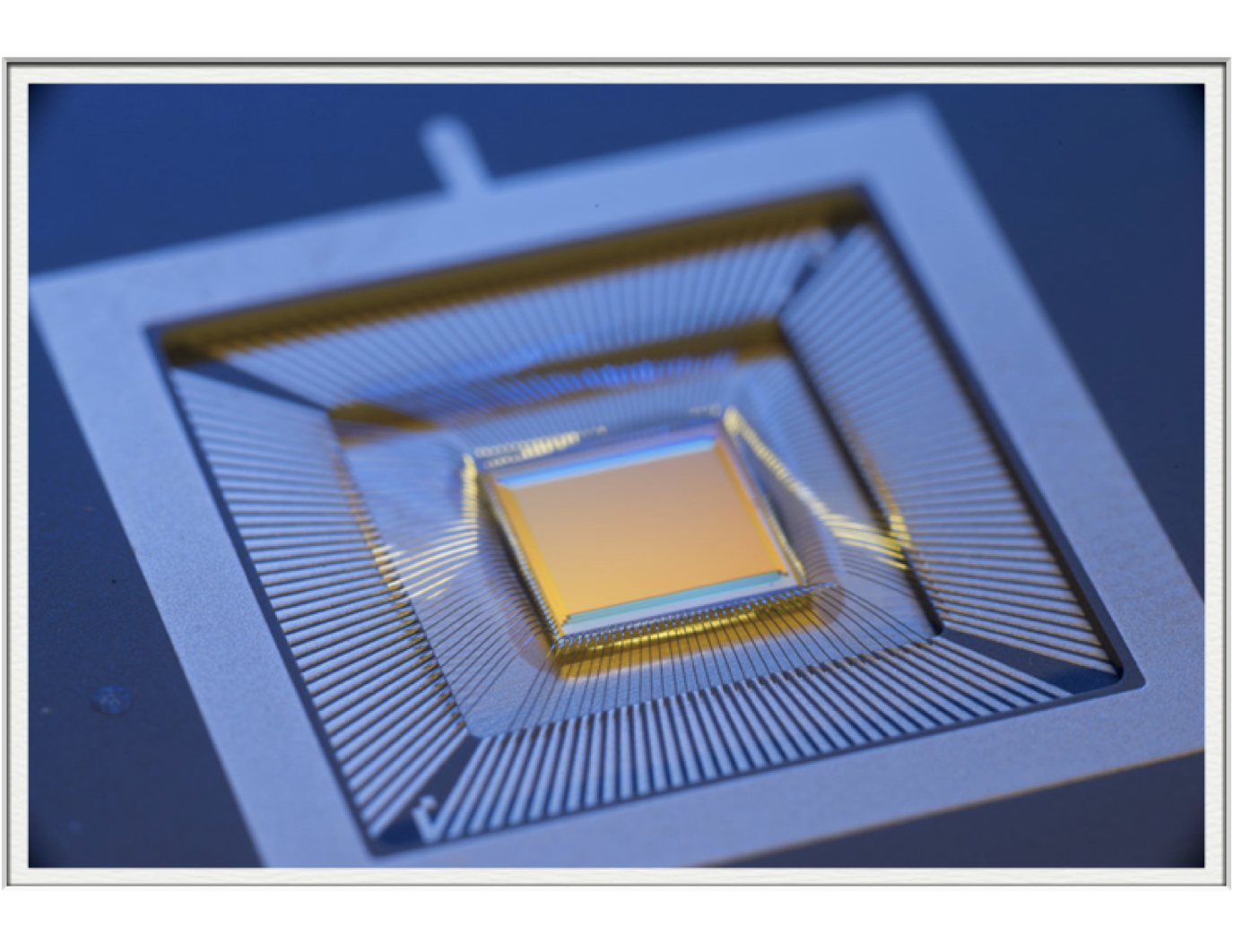}
	\caption{}
	\end{subfigure}
\caption{Block diagram of the \protovipram~(a) and picture of \protovipram~(b).}
\label{fig:protovipram}
\end{figure}

The chip operates in two modes: {\it load mode} and {\it run mode}.
In load mode, patterns are stored in each PRAM, one at a time.  
High speed performance is not required in load mode as this typically happens between running conditions with respect to the pattern matching during run mode.  
This pre-defined set of patterns is determined from offline simulation (and can be determined from real tracks in data in the future)
and is often referred to as a {\it pattern bank}.
In run mode, incoming data is compared to the stored patterns and matched pattern flags are generated based on the match threshold conditions asserted.
At the end of the event, an "event re-arm" signal can be asserted to clear all the matches and data, and the next event can be inputted.  
There is a 32-bit output which reads out the matched pattern flags, corresponding to each of the columns of a selected row.  
The readout implementation is kept simple to allow for easy testing of the pattern matching performance.
In Fig.~\ref{fig:protovipram}(a), a block diagram of the \protovipram~functionality is shown.  
In addition to the four 15-bit inputs and 32-bit output, 5(7)-bit inputs are used to designate the column (row) and there are inputs related to the clock signals and power inputs.  
There is also one more bit available for each of the four inputs which we designate as "data valid" bit.  This denotes whether or not to ignore the incoming data and allows us to invalidate certain inputs due to dead detector elements
or to handle variations in the number of inputs per layer. 

The chip has a multi-VDD design, which allowed us to study the power behavior of the chip in great detail. 
The power inputs of the chip are $V_{\rm DVDD}$, $V_{\rm DVD}$, and $V_{\rm charge}$. 
$V_{\rm DVDD}$ drives the input and clock buffers, $V_{\rm DVD}$ supplies the majority logic and the SRAM storage cells inside the CAM cells, and $V_{\rm charge}$ charges the matchline inside the CAM cells.

Figure~\ref{fig:protovipram}(b) shows a picture of the actual \protovipram~wire-bonded to a standard 144-pin Thin Pin Grid Array (PGA).   
The design has been thoroughly simulated at all levels with timing, signal dispersion, and power consumption.
Further details on design and on the simulation checks on the design can be found in~\cite{Liu:2015oca}.

\section{Test bench for performance studies}
\label{sec:testbench}

\begin{figure}[tbh]
\centering
	\begin{subfigure}[bht]{0.47\textwidth}
	\includegraphics[width=\textwidth]{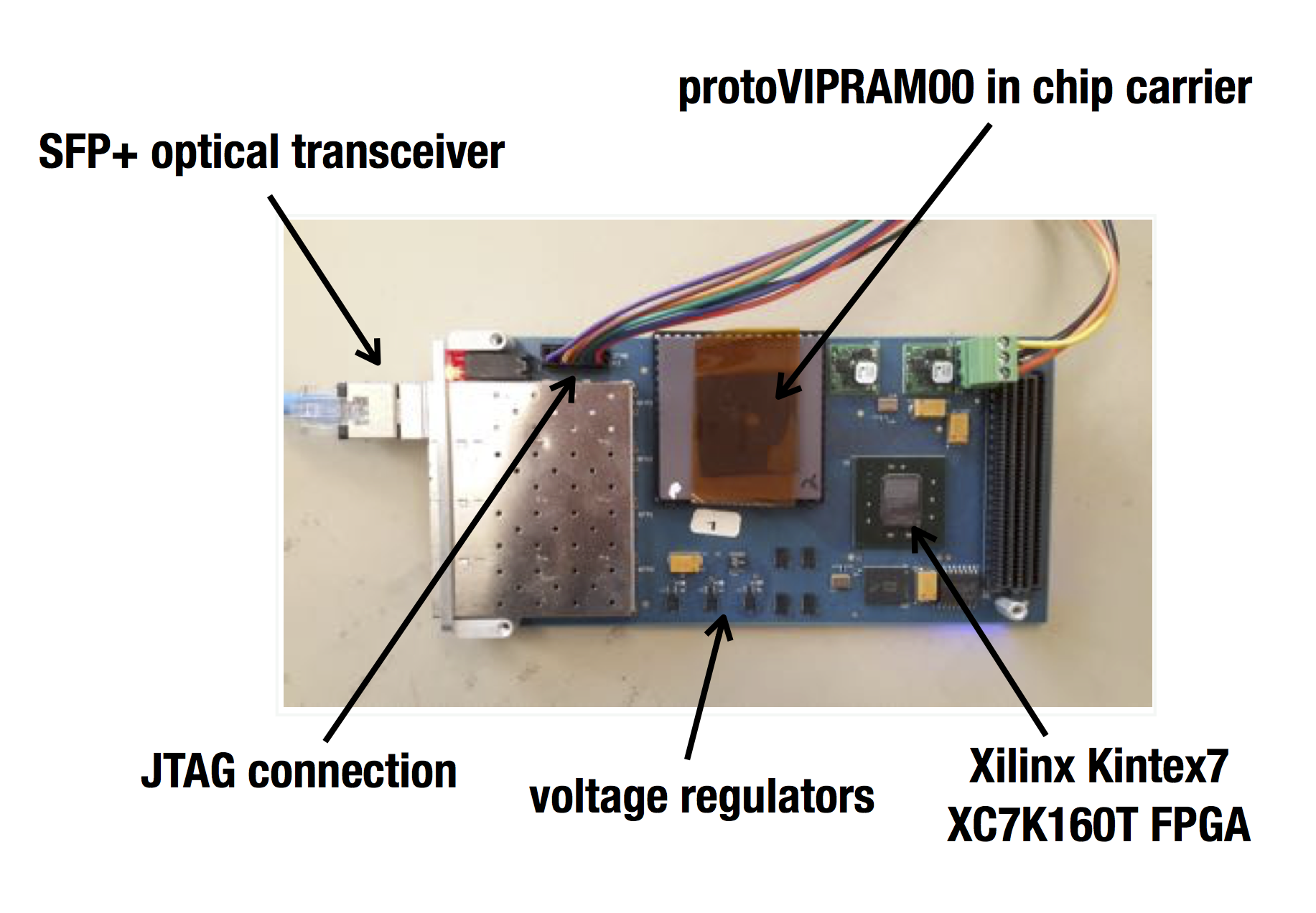}
	\caption{}
	\end{subfigure}
	\begin{subfigure}[bht]{0.47\textwidth}
	\includegraphics[width=\textwidth]{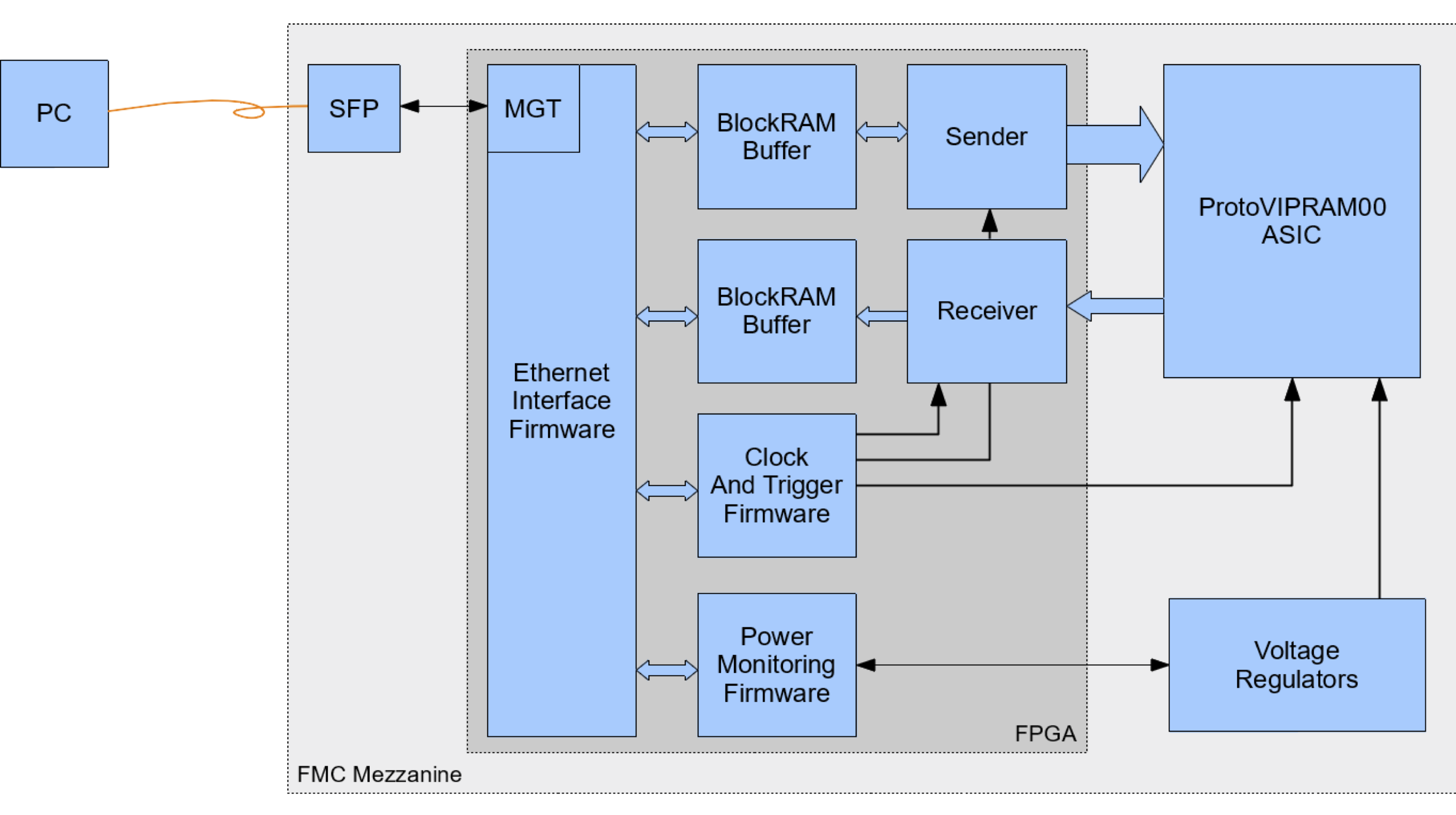}
	\caption{}
	\end{subfigure}

\caption{Mezzanine card with \protovipram~mounted (a) and block diagram of testbench data flow (b).}
\label{fig:mezz}
\end{figure}

The \protovipram~chips, wire-bonded in a socket, are mounted on a test mezzanine card, see Fig.~\ref{fig:mezz}(a). 
The FMC Test Mezzanine card features a Xilinx Kintex XC7K160T FPGA, 4 SFP+ optical transceivers, 128MB DDR3, and a 144 pin socket used for testing custom ASIC chips which are
indicated in Fig.~\ref{fig:mezz}(a).
The power supply provides 1.5~V to each of the three power inputs of the chip, $V_{\rm DVDD}$, $V_{\rm DVD}$, and $V_{\rm charge}$.
We can configure each of these power inputs separately.
The mezzanine is connected to a PC running the Linux SL6 operating system.  
JTAG Communication with the FPGA is done using Xilinx Design Suite.
Gigabit Ethernet communication is provided via the SFP+ optical transceivers. 

Testing proceeds in two ways: {\it basic functional validation} and {\it automated testing}.  
Basic functional validation stores test vector data in internal FPGA blockRAMs and the chip output is analyzed via internal FPGA logic analyzer, ChipScope Pro.
Studies are limited by the size of the FPGA memory and interpreting results from the logic analyzer output.

The automated testing data flow is illustrated in Fig.~\ref{fig:mezz}(b).
Automated testing stores test vector input and output in internal FPGA blockRAMs and reads them in and out via an optical Gigabit Ethernet connection based 
on simple UDP packet transfers (the IP-bus protocol~\cite{ipbus}).
The software needed to initialize, control, and transfer data to and from the FPGA and to analyze the data is custom written in Python.
The clock frequency is dynamically controlled in the software by dividing the internal Kintex-7 FPGA 1~GHz clock by integers.  
Therefore, allowable testable frequencies are, $f = 1000/n~{\rm MHz}$ where $n$ is an integer.
Functional tests were performed in the frequency range from 2 to 166 MHz.  
Additional current monitors are available on the mezzanine to monitor the voltage sources to \protovipram~provide measurements of the chip power consumption.
The sampling frequency of the current monitors is approximately $\sim$1~kHZ which is slower than the operational frequency 
and so when measuring power consumption we repeat functional tests serially to get a consistent current measurement. 
Measurements from the current monitors are sampled until the power consumption asymptotes so that we can obtain reliable power measurements.
The generated input patterns and the sampled outputs from the chip are verified against functional simulations done on the full chip design using Cadence NC-Sim.  

There are additional factors which can affect the performance of the chip and which should be considered in the testing.
First, because of the way that the CAM cell is designed with a 4-bit selective pre-charge, the order of the bits of the patterns injected can change the chip performance.  
Current flows through the CAM cell matchline when the 4-bit selective pre-charge is matched, so putting often-matched bits in the logic cells can consume more power in the chip overall.
Second, the power supply voltage supplied to the \protovipram~can affect the performance.
Our multi-VDD CAM chip can operate on a wide range of voltages for the three power inputs, which can be optimized to get better performance at lower power consumption~\cite{multivdd}. 
For the studies presented here, we nominally run at 1.5V. 
Details will be discussed further below.
Finally, the \protovipram~clock is supplied by two signals which we designate $MC_A$ and $MC_B$.
The clock signal is defined logically by `$MC_A$~\&\&~!$MC_B$' and the signals are offset by a phase delay.
The offset determines the discharge time of the Matchline.  If the Matchline is not given sufficient time to discharge, this would result in testing results giving "false positives", e.g. matches when no match exists.
The first two are subdominant effects though they are worth quantifying.
The final factor can result in large "false positive" matches and should be carefully considered.  
These factors will be discussed further in the testing results.

\section{Results}
\label{sec:results}

\subsection{Functional validation}

\begin{figure}[!b]
\centering
\includegraphics[width=3.5in]{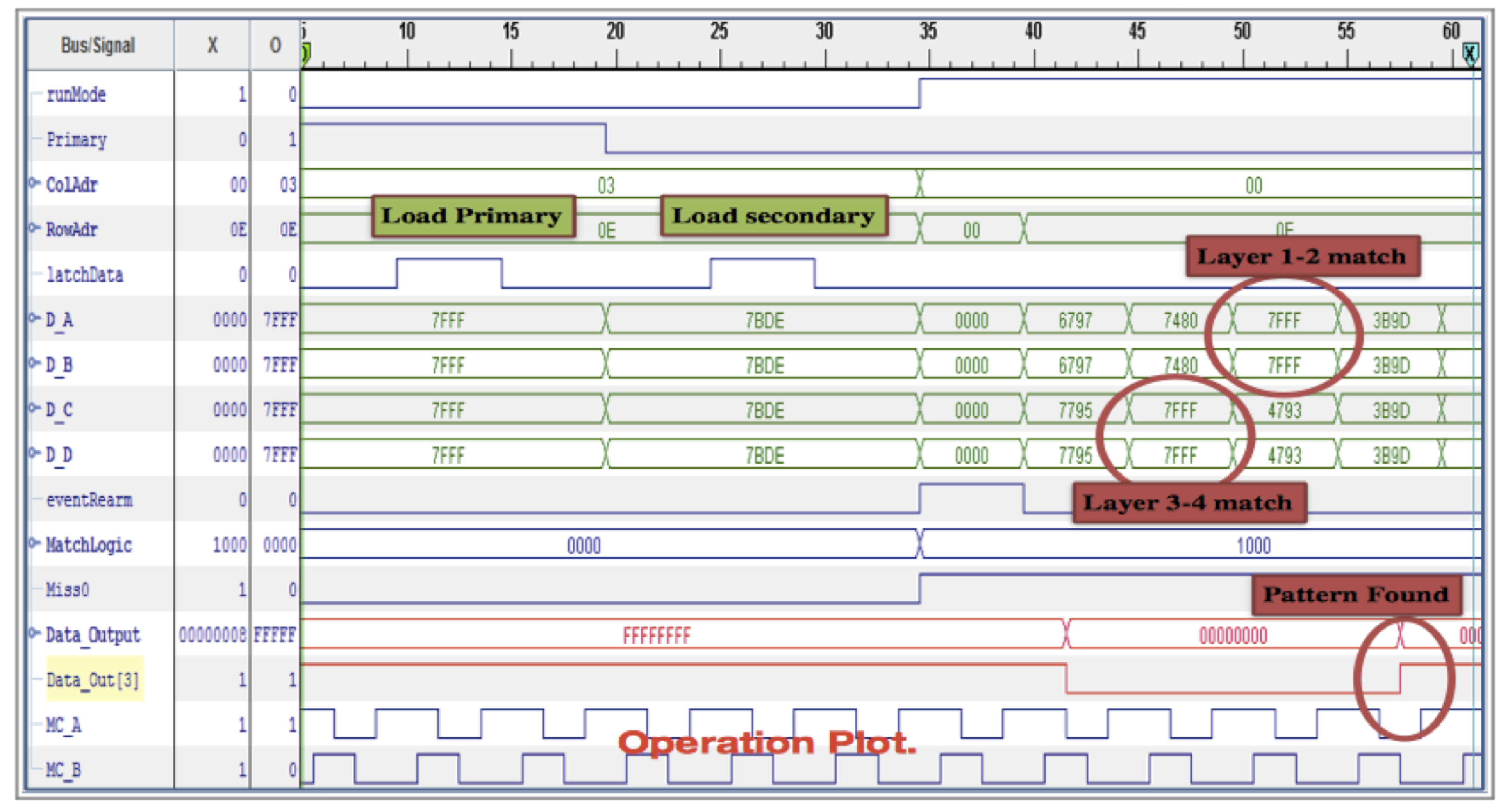}\\
\caption{\protovipram~I/O waveforms captured by the Chipscope embedded logic analyzer firmware}
\label{fig:basic}
\end{figure}

In functional testing, a simple analysis can be done using the logic analyzer output to test very basic chip functionality as a generic pattern matching device.  
An example is given in Fig.~\ref{fig:basic}.
The green text boxes indicate the clocking in of a dummy pattern into a PRAM.  
Loading a pattern into a PRAM consists of a primary load step and a secondary load step to verify the pattern.
The red text boxes indicate the searching of the memory for the input dummy pattern where the final step is the appearance of the found pattern on the output bits.
Although the level of sophistication of analysis is limited, further tests using the basic testing setup are performed including checks of each PRAM's functionality.

With the functional testing, we can verify that each pattern within a chip is working properly.  
In this test, we "walk" through each column in the chip and check that each row element in the chip can match a pattern.  
Many variations on the pattern validation are performed.  
For example, we try a number of different input patterns or varying the patterns and we also maximally vary the inputs and the patterns flipping all the bits in a pattern.
No issues in functionality are observed.    
Generally, we find that no chips have partial functionality, they are fully functional.
We check other functional behavior such as the data valid bit.  We verify that setting the data valid bit to 0 ignores the input.  
We also check the various majority logic for functionality and we find no issues with any of the possible logic states: All Layer Match, One Missing Layer Match, Two Missing Layers Match, and First Layer Match.
For the remainder of the following test, we typically stay with "All Layer Match" patterns for consistency.

Finally, during the functional testing, we lay out the timing parameters of the different chip operational modes.  
We find that for {\it load mode}, we are able to load the patterns into memory into the chip at a frequency of 10 MHz.  
Operating load mode at a higher frequency can sometimes cause a pattern to not be properly loaded.  
However, as was discussed in Sec.~\ref{sec:design}, the operational frequency of load mode is not an important parameter for performance of the chip as timing constraints are not strong when loading patterns.  
Alternatively, the operational frequency in run mode is of paramount importance. 
We study this in great detail below and understand chip performance against a number of parameters.
The remaining important timing parameter is output latency onto the output bits.  
We generally find, although a pattern match is found based on the run mode frequency, it may show up on the output bus with a typical delay of 5~ns.  
We hypothesize that this comes from pushing a signal through the low speed chip carrier PGA package although the final configuration is for the chip to be bonded directly to the mezzanine.
The chip is not designed for fast readout and we study this in greater detail for future chips, as mentioned above, and we note this simply as an issue when designing our tests.

\subsection{Realistic HL-LHC scenarios}

After verifying functional performance of the \protovipram, we go on to test the chip performance using realistic HL-LHC scenarios.
Automated testing allows us to perform large scale tests of the entire chip, both by providing test vectors larger than 
the size of the internal FPGA RAM and analyzing output with ASCII output,
without having the re-program the FPGA.  
The data flow is from the custom software through SFP+ optical transceiver to the Kintex7 FPGA 
from which the test vectors are pushed to the \protovipram~where the output is written back to the FPGA and then back to the software package.
We are able to scan in operational frequency and monitor chip power consumption in real-time.
Input vectors are created and sent via UDP packet transfers to the blockRAMs which are then sent to the chip.  
Output signals are captured and stored in blockRAM to be read by the analysis software.
This is done iteratively if the size of the test vectors exceeds the blockRAM of the FPGA, 
which fits $2 \times 10^{15}$ clock cycles of test vectors.  

We use simulated data from high occupancy HL-LHC collisions and input the data as if it was coming from the collisions to the triggering system. 
As we introduced in Sec.~\ref{sec:intro}, the associative memory technique is used for identifying charged particle tracks in the detector, creating a set of known detector patterns.  
The challenge is to identify these patterns with an extremely low latency in with a lot of noise (uninteresting hits) from additional low energy collisions.
Therefore, in order to benchmark the \protovipram~performance in realistic scenarios, we use simulation of from the CMS experiment in the HL-LHC collision environments. 
The simulation provides for us a set of allowed patterns in the detector ({\it pattern bank}).
Of course, we are not yet testing results for an entire detector which requires millions of patterns.  
The inputs for our realistic tests are defined as a unique set of patterns taken as a subset of a full pattern bank from the CMS experiment 
HL-LHC simulation.
A {\it trigger tower} is defined as a regional partitioning of the entirety of the detector hits into various detector regions; 
for example, we simulate the CMS detector split into 48 trigger towers.  
In our realistic tests, we emulate a part of one such trigger tower.  
We note that the parameters below we use for testing are not near the requirements for the final system but instead simply a test benchmark based on our best knowledge at the time.

The set of allowed hits are then as the hits within our mock pattern bank which is part of a trigger tower.  
The data from the HL-LHC detector comes as charged particle hits on the various layers of the detector.  
The hits on the various layers come randomly ordered.
The number of hits per layer is taken from HL-LHC simulation of the CMS detector with high pileup and four top quark events.  
From simulation studies, we take the benchmarks for the average number of hits per layer as:

\begin{itemize}
\item Layer 1: 90 hits
\item Layer 2: 60 hits
\item Layer 3: 45 hits
\item Layer 4: 35 hits
\end{itemize}

From those sets of hits, we randomize them and send them to the \protovipram.  
Additionally, from simulation, we determine that a typical event has 5 true tracks for our mock trigger system.
The hits of those true tracks are also mixed in with the random hits and sent to the chip.  
The threshold for success for a given event is to find the patterns for those 5 true tracks. 
Note that the fraction of true matches is extremely small ($\sim5/4096$) with respect to the number of total patterns in realistic HL-LHC scenarios.
We define the matching efficiency for real tracks in these realistic tests as: $\varepsilon_{\rm match}  = N_{\rm found}/N_{\rm expected}$.
We also look for {\it false positives} which are fake matches despite no expected real track, and we find the contribution from false positives 
in our current settings, to be negligible.  
We discuss this more below.

Results of realistic testing are shown in Table~\ref{tab:table1} as the matching efficiency as a function of the operational frequency, the speed 
at which we introduce test vectors to the chip and search for patterns.  
The \protovipram~shows 100\% matching efficiency up to the target operational frequency of 100~MHz, which was the design goal of the prototype chip.
Further optimization of the pattern bank can improve the performance of the chip.  
If we have a full match of the 4-bit selective pre-charge for a given CAM cell, then power is driven through that cell.  
By distributing evenly how likely the 4-bit selective pre-charge is matched across the entire pattern bank, we can reduce power consumption across the chip and moderately improve the operational frequency
of the chip.  
This can be seen by comparing the second and third columns of Table~\ref{tab:table1}, where "re-ordered" is the optimized pattern bank.
We also verify two other factors: the effect of input supply voltages and the effect of clock phase on the Matchline discharge time. 
We vary the voltages supplied to $V_{\rm DVDD}$, $V_{\rm VDD}$, and $V_{\rm charge}$ from 1.4V to 1.6V. 
$V_{\rm DVDD}$ and $V_{\rm VDD}$ have a very small effect on the performance of the chip, which is dominated by the $V_{\rm charge}$ supply. 
We run with the default recommended settings of 1.5~V although we find that scanning through each of the other voltages could yield approximately a 10\% improvement in operational frequency.
Details on the effects and optimization of the multi-VDD supplies on CAMs and the power modeling can be found in~\cite{Li:ICCD2015,multivdd}.
We also vary the phase of the two clock signals.  By decreasing the time per clock cycle devoted to discharging the Matchline, we increase the available time for doing pattern matching.
We ultimately find that the discharge time must be $\geq 1~{\rm ns}$ or else we begin to observe false positives in our testing results.

\begin{table}[h!]
  \centering
  \begin{tabular}{|c|c|c|}
    \hline
    Freq (MHz) & $\varepsilon_{\rm match}$ (default) &   $\varepsilon_{\rm match}$ (re-ordered) \\
    \hline
    \hline
    50 & 100\%  & 100\%     \\
    \hline
   60 & 100\% & 100\%  \\    
    \hline
    71 & 100\%    & 100\% \\ 
    \hline
    76 & 100\% & 100\%     \\
    \hline
    83 & 100\% & 100\%     \\
    \hline
    90 & 100\%  & 100\%    \\
    \hline
    100 & 100\% & 100\%     \\
    \hline
    111 & 99.76\% & 100\%     \\
    \hline
    125 & 95.3\%   & 98.8\%   \\
    \hline    
     \hline
  \end{tabular}
    \vspace{0.25cm}
  \caption{Matching efficiency of the \protovipram~chip as a function of operational frequency. }
 \label{tab:table1}  
\end{table}
We perform realistic tests of 12 wire-bonded \protovipram~chips and show consistent performance across all chips tested.

\subsection{Extreme boundary conditions}

\begin{figure}[!t]
\centering
\includegraphics[width=3.6in]{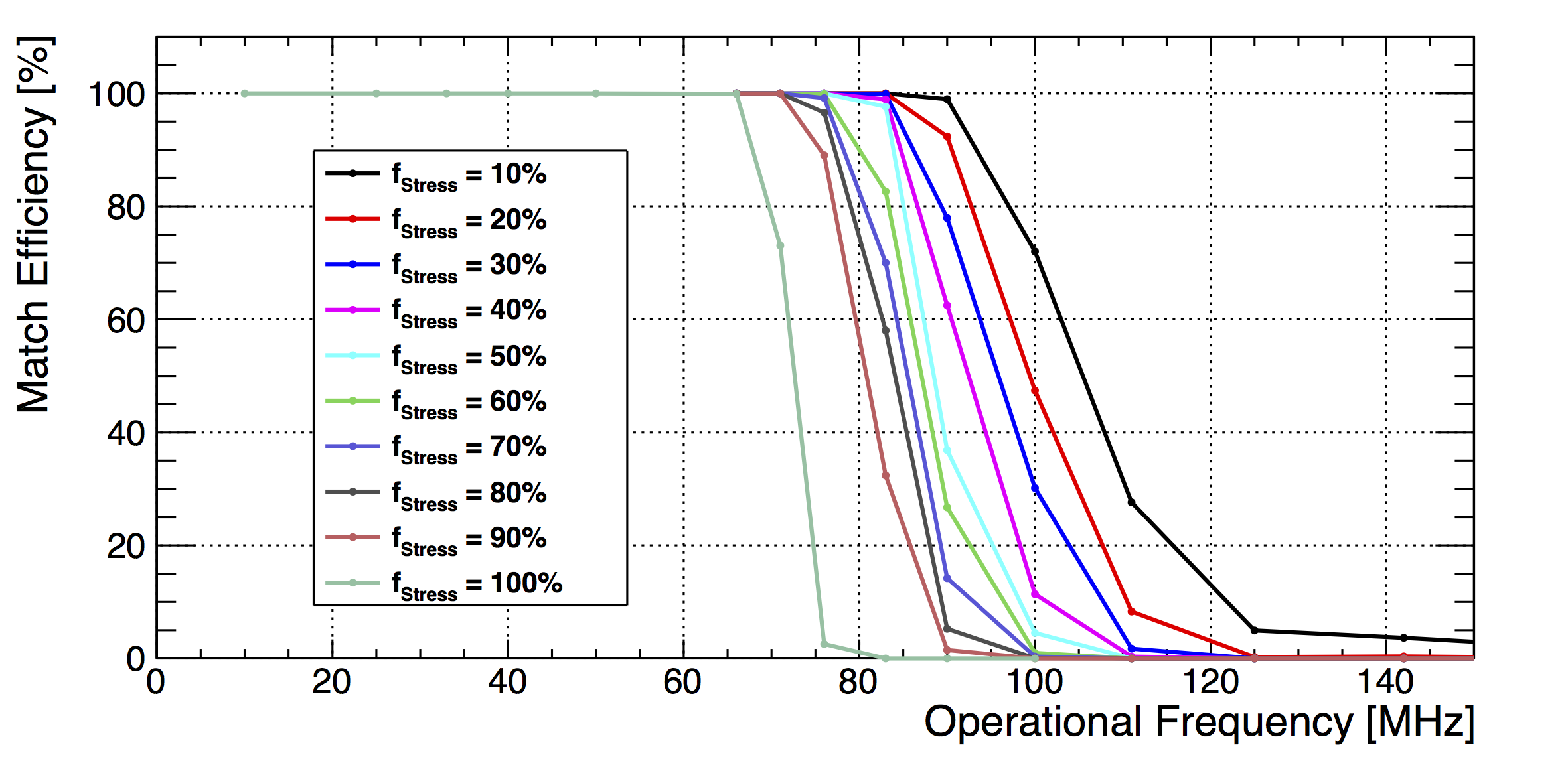}\\
\caption{Results of extreme boundary condition testing on \protovipram}
\label{fig:stress}
\end{figure}

To understand the bounds of the chip performance, we use dummy data to test the \protovipram~in scenarios far exceeding what we expect in realistic scenarios.  
These extreme boundary condition tests are used to benchmark the limits of the chip performance and understand systematically its limitations and breakdown points in terms of match occupancy and operational frequency.  
Additionally, performing a detailed study of the power consumption with these tests will guide us to finding any improvements for future chip designs.
Many tests are performed to test the performance of the \protovipram~and here we describe the most complete set of tests.  
The typical match occupancy for a realistic system is $< 1\%$ of patterns matching within a given event and matches do not happen simultaneously in time because
the hits arrive at the chip randomly and matches occur throughout the entire event.  
In these extreme boundary condition tests we force matches within a given event to occur in the same clock cycle, e.g. there is only one clock cycle in the event.  
Further, we require $\gg1\%$ of the chip to match at the same time, scanning a fraction, $f_{\rm ext}$, of 10\% to 100\% of the chip simultaneously expected to match. 
Here, the subscript ``ext" refers to the extreme fractional occupancy of the chip. 
We do this by filling the initial pattern bank with only 2 unique patterns occupying $f_{\rm ext}$ and $1 - f_{\rm ext}$ of the chip, respectively.  
They are distributed evenly through the chip geometrically in order to not bias the tests based on the location of the patterns in the chip layout\footnote{Other tests were made to determine which location of the chip was most likely to fail with a moderate preference for the middle of the chip where signals take the longest to propagate.}.
We then send the pattern which constitutes $f_{\rm ext}$ of the pattern bank and check to see how many of them matched.  
This test is performed many times to get a large set of statistics from which to compute the matching efficiency.
We perform this test at various frequencies and determine when the chip performance begins to degrade.  
The results of the extreme boundary condition tests can be seen in Fig.~\ref{fig:stress}.

One can see that as the operational frequency of the chip is increased, the performance begins to degrade.  
The frequency at which the performance begins to degrades decreases as we increase $f_{\rm ext}$.
This behavior is expected.  
As the operational frequency increases the majority logic voltage has less time to give a matched signal.  
As the fraction of the chip that is matched increases, $f_{\rm ext}$, the signal propagation throughout the chip becomes more delayed with increasing matches
and the chip begins to miss expected matches.  
The testing results quantify where chip degradation occurs and where the chip ultimately can still perform in the most extreme scenarios.

\subsection{Power Consumption}

CAMs are very attractive for pattern recognition applications due to their high speed performance, however they incur significant power and area overheads.  
The primary reason for this is the massively parallel operation of the CAMs.
The massively parallel structure in a CAM needs a large amount of driver circuits to multiply and drive the input data signal to each of the CAM cells.
Therefore, it is important to measure the power consumption of the chip and properly model it so that we can estimate requirements for larger scale systems.  
The multi-VDD design of our chip, with a separate supply for the major functional blocks, allowed us to study the power behavior of the chip in detail and model it.
This will help us to understand the absolute scale of power consumption for an
ultimate track trigger system; validate the breakdown in performance of the chip and extrapolate its performance to 3D.
Recall from Sec.~\ref{sec:fullchip} that the power inputs are $V_{\rm DVDD}$, $V_{\rm VDD}$, and $V_{\rm charge}$ where
$V_{\rm DVDD}$ drives the input signals, 
$V_{\rm VDD}$ primarily drives the majority logic,
and  $V_{\rm charge}$ charges the Matchline. 

Programmable voltage regulators on the mezzanine card support current readback for the chip.
The dominant source of power consumption in \protovipram~comes from the input line drivers, $V_{\rm DVDD}$,
and for an operational frequency of 100 MHz, the typical power consumption of the chip is approximately 250~mW.  
Scaling the power consumption of the pattern matching is non-trivial for the 4096 pattern chip but in the given prototype it is not the dominant contribution.

\begin{figure}[!tb]
\centering
\includegraphics[width=3.6in]{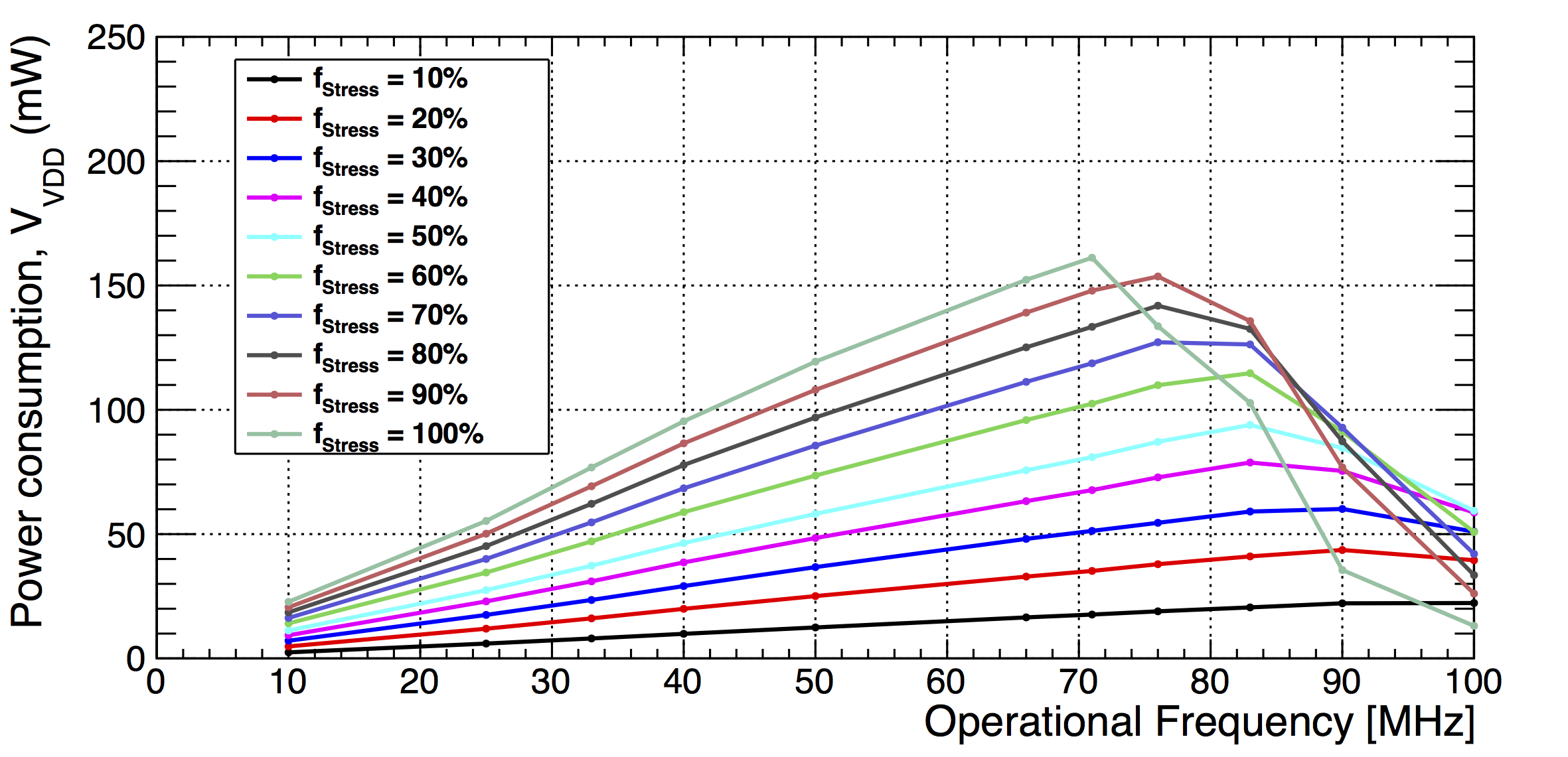}\\
\caption{Power results of stress testing on \protovipram}
\label{fig:pwerstress}
\end{figure}

In unrealistic extreme boundary conditions, we monitor all power lines and can use it to verify the performance of the chip. 
For example, we can monitor $V_{\rm VDD}$ in our extreme boundary condition tests. 
This is shown in Fig.~\ref{fig:pwerstress}.
The $V_{\rm VDD}$ shows the matched pattern power consumption and as we increase $f_{\rm ext}$ and the operational frequency and the chip performance
begins to degrade, we can see that the power of $V_{\rm VDD}$ also correspondingly begins to degrade.  
This provides us with an excellent validation of our understanding of the internal chip functionality.

Beyond measurements of the power consumption, a methodology for characterizing the power consumption of the chip has also been developed. 
It shows excellent agreement with the testing results.  
Chip behavior can be well-modeled for a variety of different factors including varying the types of patterns and selective pre-charge as well as the dependence on the voltages.  
Overall, our model can predict the average total power consumption of the chip to within 4\% of the actual measurement.  
Many more details about the modeling of the chip power consumption and results can be found in ~\cite{Li:ICCD2015,multivdd}.


\section{Conclusion and Outlook}
\label{sec:concl}

Fast triggering on particle tracks is vital to the physics program of the HL-LHC at CERN.
Associative Memory-based pattern recognition provides a powerful approach to solve the complex combinatorics inherent to this challenge.
The PRAM (pattern recognition associate memory) devices that are at the core of its concept are well-suited to modern 3D integration. 
Emerging 3D technologies provide an opportunity to improve in pattern density while simultaneously improving speed capabilities and reducing power consumption.
The first \protovipram~chip was designed and fabricated in a 130 nm Low Power CMOS process. 
The layout was deliberately implemented in 2D so that the basic associative memory building blocks can be directly re-used for 3D stacking. 
The design has been successfully tested both for functionality and performance using a custom test setup.
Results indicate that the chip operates at the design frequency of 100 MHz with perfect correctness in realistic conditions and 
conclude that the building blocks are ready for 3D stacking.

The set of testing results above span the possible parameters of the chip we could conceive.  
We checked all the basic functionalities of the chip outlining generic performance parameters.
Then we performed tests using realistic HL-LHC scenarios and also devised scenarios to understand the boundary conditions of the \protovipram.  
We varied different parameters which could affect the chip performance from the clock phase, input voltages, pattern order, and operational frequency.
We tested a number of chips and found consistent performance across the set.
In all cases, we find the chip to behave within design specifications and gained a large amount of experience in how to test the chip, troubleshooting various issues, 
and defining a baseline for the type of tests that should be performed when studying future AM chips.
The testing results show that the basic associative memory building blocks, 
the CAM and the control cell that comprise \protovipram~are ready for 3D vertical integration for a proof-of-principle demonstration of the VIPRAM concept. 

Following successful performance evaluation of the \protovipram~presented in this paper, two new chips have been developed to split the development path towards different goals.
The {\tt VIPRAM\_3D}~\cite{Liu:2011zzw} is meant as a verification of multi-tier 3D stacking and is identical to the \protovipram~in all relevant design choices except that it will be stacked in 3D. 
It is fully pin-compatible with the 2D \protovipram~and can provide a direct diagnostic of the 3D integration process.
The {\tt VIPRAM\_L1CMS}~\cite{vipram_l1cms} focuses on bringing the system interface to maturity including pipelined operation and sparsified readout and includes 3D integration of 2 layers: the IO tier and the PRAM tier. 
The design of the next generation VIPRAM chips have been completed, the wafers have been fabricated, and they are now in 3D processing.


\section*{Acknowledgment}
Operated by Fermi Research Alliance, LLC under Contract No. De-AC02-07CH11359 with the United States Department of Energy.

\ifCLASSOPTIONcaptionsoff
  \newpage
\fi



%




\end{document}